# A Preliminary Exploration on Component Based Software Engineering

N Md Jubair Basha[†], Dr Gopinath Ganapathy[††], and Dr Mohammed Moulana[†††]

[†]Research Scholar, Department of Computer Science, Bharathidasan University, Tiruchirapalli, Tamil Nadu, India
[††]Professor & Head, Department of Computer Science, Bharathidasan University, Tiruchirapalli, Tamil Nadu, India
[†††]Professor, CSE Department, Koneru Lakshmaiah Education Foundation, Guntur, Andhra Pradesh, India

**Abstract**
Component-based software development (CBD) is a methodology that has been embraced by the software industry to accelerate development, save costs and timelines, minimize testing requirements, and boost quality and output. Compared to the conventional software development approach, this led to the system's development being completed more quickly. By choosing components, identifying systems, and evaluating those systems, CBSE contributes significantly to the software development process. The objective of CBSE is to codify and standardize all disciplines that support CBD-related operations. Analysis of the comparison between component-based and scripting technologies reveals that, in terms of qualitative performance, component-based technologies scale more effectively. Further study and application of CBSE are directly related to the CBD approach's success. This paper explores the introductory concepts and comparative analysis related to component-based software engineering which have been around for a while, but proper adaption of CBSE are still lacking issues are also focused.

## 1. Introduction

A well-known subfield of software engineering is component-based software engineering. Object-oriented design, software architectures, architecture definition languages (ADLs), middleware, and various development methodologies like structural and modular development are the sources of the techniques and technologies that make up CBSE. The idea of creating systems out of parts originated in other engineering fields.

The issue of locating a standard and sufficiently accurate definition of the phrase "software component" presented itself to CBSE right away. Software components are compositional entities having explicitly stated context dependencies and interfaces that are legally determined. A software component may be independently offered and is open to third-party compilation[1].

The creation of component-based software is essential for boosting a business' productivity. A wide variety of reusable components are required in the repository. Once the requirements are documented, the development activity typically begins from scratch. Overruns in both time and money may occur from this. Instead of starting from scratch and creating the entire system, it is faster and more effective to reuse an existing component. Setting up a software recycling process requires a primary asset, however only few recycling operations make use of this asset.

In other words, building a reuse process and repository establishes a knowledge base that raises the caliber of the final result after each cycle of reuse. This lowers the risk of new initiatives based on repository knowledge and lowers the development work for subsequent projects. As the component will already have been successfully tested in the repository, it also aids in reducing the testing work.

The introduction to component-based software engineering and the many stages of its implementation in software development are covered in this paper. The following sections make up the remainder of the paper. Section II compares CBSD with traditional software development to outline the fundamental ideas of CBSE. The CBSD lifespan is presented in Section III. The various scripting technologies needed for CBSD are covered in Section IV. The technologies based on components are presented in Section V. Section VI compares the qualitative performance of scripting and component-based technology. Section VII outlines the issues and obstacles that still need to be overcome to complete various works in component-based software engineering.

## 2. Basic Concepts of CBSE

The CBSE model suggests separating component creation from system development and creating systems out of reusable components. This split has important effects on economic goals, such as creating a market for components, technological developments, such as offering new functionality immediately, as well as legal and societal concerns (e.g. trust, accountability and maintenance). The CBSE is based on the following four guiding principles to achieve its main objectives of improved development efficiency, quality, and shorter time to market.

**2.1 Reusability:** Only if the components can be utilised again in various applications after they are constructed will the CBSE strategy be effectively adopted. Commercial off-the-shelf (COTS), product line, and open source components are only a few examples of reusability types that the industry has recognised as best practices. When developing architectural components for a particular system with no plans to reuse them in other systems, CBSE is also helpful.

**2.2 Substitutability:** Systems stay proper even when a component is substituted thanks to substitutability. The Liskov substitution principle is what this stipulation amounts to

Let q(x) be a property that can be demonstrated for all x-type T objects. If S is a subtype of T, then q(y) should be true for objects y of type S. For functional features, this idea is reasonable, but it is less clear for extra functional properties because it depends on other aspects, such, for instance. B is subject to the system context. A speedier component, for instance, could deadlock and interfere with timing demands in a system that employs a non-preemptive scheduling technique.

**2.3 Extensibility:** By incorporating new components or enhancing already existing ones to improve system functioning, extensibility in CBSE attempts to enable evolution. Giving components numerous interfaces is a common way to support component evolvability.

**2.4 Composability:** Composability is a cornerstone of CBSE education. The composition of functional qualities is supported by all component-based technologies (component binding). Less frequently, the composition of non-functional qualities, such as component dependability, execution time, or memory use, is supported. One of the primary issues facing CBSE research is the assembling of extra functional characteristics.

The following are the primary advantages of employing the software components[2]: Cost and development time savings: Because the component is reused, these factors are reduced. The component can be changed if necessary. Reduced Testing Effort: Since testing takes up more than 60% of the time spent developing software. The domain-specific component approach reduces test effort. A higher level of quality has been achieved because every successfully developed component is certified. As a result, the components in the repository are frequently of high quality.

Many businesses have created their own domain-specific parts that serve as assets and can be used again in the future. The component can be changed even if it is not mirrored and reused. A component change requires less work than a new development. However, a practical method for locating and creating the components is required. The distinction between conventional and component-based software development is described here.

Table 1: Comparison of CBSD and Traditional Software Development [3]

| Component Based Software Development | Traditional Software Development |
|---|---|
| Housing System from already available components. | Housing system from bottom-line. |
| Components and Systems integrated from those components are developed through interfaces. | Software System is developed by following all the phases of life cycle. |
| Component selection, identification and evaluation are special life cycle phases. | There is no provision for selection, identification and evaluation in this life cycle. |
| Effort is required for component selection, testing & verification only once. | Much effort is required for throughout the software development cycle. |
| Reuse of components guide to development of component in a faster manner | Reuse property is applicable in a less manner. |
| Cost and time management is required less. | Cost and time management is applicable for every project. |
| Depends on the requirements, component management can be done applicable to the project. | Software development activity has to be carried out for every project. |

## 3. Component Based Development Life Cycle

The component-based development model incorporates the characteristics of spiral models (CBD). The applications of the grouped software components make up the CBD model (called classes). The identification of potential components serves as the first step in the component development process. This is accomplished by determining the data that the programme will modify and the pertinent algorithms that will be used. The classes serve as a container for the data and algorithms.

The component-based development model incorporates the characteristics of spiral models (CBD). The applications of the grouped software components make up the CBD model (called classes). The identification of potential components serves as the first step in the component development process. This is accomplished by determining the data that the programme will modify and the pertinent algorithms that will be used. The classes serve as a container for the data and algorithms. The repository houses the software project components that were produced. The repository is mined to see if the desired components are present after candidate components have been discovered. They are collected and reused when they are accessible. If the component is created using object-oriented technique and does not already exist in the repository. The components for the application's initial build are taken from the repository and new ones are created to accommodate the changing requirements of each application.

As demonstrated in Figure 1, the process flow eventually loops back to the spiral model and continues the component assembly loop over successive iterations of the component lifecycle. The CBD model can be used to achieve software reusability, which is particularly beneficial for software engineers. Software reuse, according to QSM Associates Inc., leads in a shorter development lifecycle, an 84 percent decrease in project expenses, and a productivity index of 26.2 as opposed to the industry average of 16.9. Software developers can benefit greatly from CBD models and component robustness repositories.

Figure 1 illustrates the sequential procedure that Sommerville [4] described for CBSD. There are six steps in total, and they are as follows:
1. Because specific needs restrict the amount of components that can be employed, the user requirements are established in broad strokes rather than in fine detail.
2. As many components as possible are found to be reusable using the given requirements.
3. The standards are stringent and specifically designed to be satisfied by the components.
4. Using the aforementioned procedures, architectural design development is achievable.
5. The architecture can be designed using this system. Repeat steps 2 and 3 as necessary.

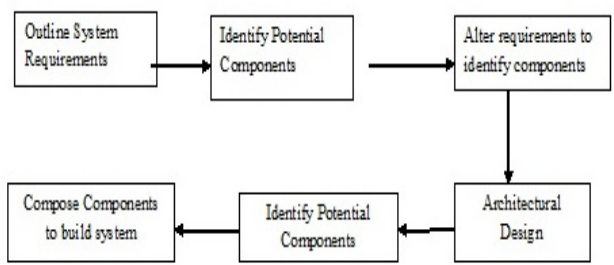

Figure 1: CBSD Process

## 4. Scripting Technologies

The foundation of scripting technologies is a language interpreter included with web server software. The interpreter normally accesses the database while processing the code that is embedded in the HTML pages. The output of the script is substituted for the script's code, and the resulting HTML code is sent back to the client. The static HTML code, often known as the HTML template, does not change. Language processors like PHP, ASP, and ColdFusion are examples of scripting technology. Because they are closely tied to the web server, scripting technologies are effective for producing dynamic content. They are perfect for monolithic, medium-sized programmes that need an effective execution environment. Large volumes of static HTML from templates with only a (relatively) tiny amount of dynamically created data embedded characterize other scripting-enabled applications.

As an illustration, consider the typical product description page of an e-commerce application, which uses an HTML template with variable data pulled from the database. On the other side, scripting languages are known for their intimate connectivity between the frontend and middle layer. It follows that great scalability is necessary for the use of web-related applications. However, scripting technologies sometimes lack built-in high-level support for the coordination and synchronisation of processes operating on several nodes, making the addition of nodes potentially necessary to achieve scalability. The majority of the widely used scripting languages come with function libraries that can be used to build this feature. However, this necessitates additional, significant programming work. Because of this, distributed web-based systems rarely use scripting technologies.

## 5. Component-Based Technologies

Software objects are used in component-based technologies to implement the application logic. Containers are unique execution contexts that are used to create these things. Java2 Enterprise Edition (J2EE), which comprises standards for Java Servlets, Java Server Pages (JSPs), and Enterprise Java Beans, is a well-known component-based technology for creating dynamic online resources (EJBs). Java classes called servlets implement a web application's application logic. They are created in a servlet container with a web server interface, such as Tomcat. Java servlets' object-oriented design promotes superior modularity, and their ability to execute different containers on various nodes enables a level of system scalability that scripting solutions are unable to provide.

The J2EE framework's building pieces are represented by Java Servlets. In actuality, they simply offer the basic methods for handling dynamic queries. Numerous elements must be taken care of by the programmer, including creating the HTML document template and managing communication with outside information sources. Due to these factors, J2EE technologies like JSP and EJB are frequently combined with Java servlets. Java Servlet API standard extensions known as JSPs enable the embedding of Java code in HTML documents. For future requests, each JSP is automatically transformed into a Java servlet. JSP pages aim to preserve the advantages of Java servlets without penalising web pages with a high percentage of static HTML templates and a low percentage of dynamic content. JSP is therefore a more effective method for creating dynamic content than Java servlets, which are better suited for handling client requests and data processing. JSP is frequently used as the default option for creating dynamic, component-based content.

EJBs are Java-based server-side software elements that allow for the creation of dynamic content. Similar to the Java servlet container, an EJB runs in a unique environment called an EJB container. Atomic transactions are supported natively by EJB and are helpful for preserving data consistency through commit and rollback procedures. Additionally, they handle persistent data across numerous queries. The overhead of these extra features causes performance to suffer. Only services that want user session persistence between various user requests to the same application should make advantage of them. Database transactions and shopping cart services in e-commerce applications are common examples.

## 6. Technology Comparison

Scripting and component-based solutions exhibit an intriguing performance. In this study, a straightforward e-commerce application is implemented using PHP scripting technology and compared to Java Servlets and EJB. In comparison to other component-based technologies, PHP offers superior performance while using the same hardware architecture. When compared to Java Servlets, the information gain is 30%; when compared to EJB, it is more than double. On the other side, Java servlets outperform script technology if the system platform has a large enough number of nodes.

Figure 2 compares the performance of two software systems on a qualitative level by examining system throughput in relation to client traffic volume. This graph demonstrates how component-based technology tends to provide the highest throughput for scripting technologies. This is influenced by their environment for more effective performance. Component-based solutions, on the other hand, scale better than scripting technologies and can achieve even higher throughput, but they often perform badly for small to medium-sized web applications. The primary driving force is its high degree of modularity, which enables the distribution of application functionality over numerous nodes.

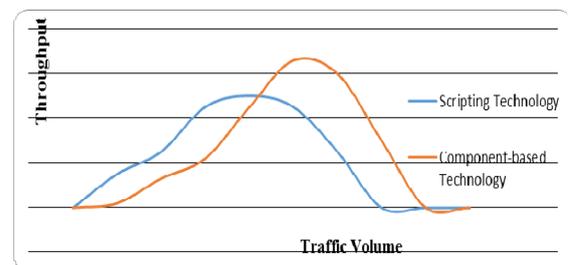

Figure 2: Qualitative Performance of Scripting and Component based Technology

## 7. Issues and Challenges of CBSE

The following are the issues and difficulties with component-based software engineering:

**7.1 Component Certification Process:** It is necessary to certify the components before classifying them. A component's certification aids in the identification of possible components. The certified components guarantee that the associated component has already been planned, carried out, tested, and used for a certain purpose. Even though certification is a common practice in many industries, it is typically not applicable to software components.

**7.2 Component Predictability:** The process by which these attributes determine the corresponding attributes of the systems that they are a part of is unknown, despite the fact that it is presumed that all pertinent components are described. Research is continuously being done to determine the best method for determining system attributes from component attributes. Is it even feasible to derive in this way? Or is it better to focus on measuring the characteristics of component composites? [1]

**7.3 Requirements Management and Component Selection:** The overall lack of completeness, accuracy, and consistency in requirements is a challenge for requirements management. The basic goal is to design a system that, within a given framework of various boundary conditions, as closely as possible complies with the requirements. Reusing previously created components is the fundamental strategy of component-based development. As a result of the potential components typically not having one or more features that completely match the system requirements, the requirements creation process is much more difficult. Even if a component fits the system well on its own, this does not necessarily mean that it will function poorly or not at all when combined with other components. Due to these restrictions, requirements engineering may need to be approached differently, the practicality of requirements in connection to the components at hand evaluated, and requirements modified as a result. A risk management approach is necessary in the component selection and development process since there are numerous unknowns in the component selection process.

**7.4 Long-term Management of Component-based Systems:** The subsystems and components of component-based systems have independent life cycles. The accumulation of components, subsystems, and autonomous life cycles makes system progression challenging. There are many various types of research questions, including technical, administrative, organizational, and legal problems. The technological considerations centered on the notion that the system could technically be updated by swapping out individual parts. Updateable components that should or must be updated make up the administrative and organizational difficulties. the legal concerns examining the system or component manufacturer's liability for a system failure. The maintainability of such systems is still carefully practiced, notwithstanding CBSE's modern approach. There is a chance that many of these systems will be difficult to maintain.

**7.5 Component Development Models:** The current development models demonstrate dominant technology, but they are difficult to use and contain a number of ambiguous qualities. The overview of the classification of component models for the software lifecycle dimension is provided by I. Crnkovic et al. in their publication [5].

**7.6 Component Configurations:** Numerous components, each of which has further components, can make up complex systems. Compositions of components will frequently be handled as components. The issues related to structure configuration suddenly appear. The same component, for instance, might be present in two compositions. Will these components be recognized as two separate entities or will they be regarded as one single identical entity? Was one of the study questions that was posed? Which version will be chosen if these components are of different versions? If these versions are incompatible, what happens? Although the issues with dynamic updating of components are already understood, their solutions are still being investigated.

**7.7 Dependable Systems and CBSE:** It is particularly difficult to apply CBD in real-time systems, process control systems, safety-critical sectors, and other systems where dependability standards are more strict. The inability to guarantee some system properties as well as the limited ability to verify component quality and other non-functional attributes is a significant issue with CBD.

**7.8 Component Quality Service:** The quality of the component's service is given more consideration at CBSE. There is still a challenge with the research subject of how Quality of Service (QoS) of components may be defined and described [6]. It is possible to see this research question as a research problem.

**7.9 Tool Support:** Component testing tools, component repositories and tools for managing the repositories, component-based design tools, runtime system analysis tools, component configuration tools, etc. are all urgently needed. Efficiently constructing systems from components is the aim of CBSE. The only way to do this is with complete tool support.

## 8. Conclusion:

Software development has been considerably enhanced by implementing a component-based software engineering method. However, there needs to be a way to find and make the components. This paper provides an introductory comparative analysis related to component-based software engineering, with the goal of formalizing and normalizing CBD-related behavior across all disciplines supported by CBSE. Therefore, component-based technologies beat scripting-based ones in terms of qualitative performance. The life cycle dimension of classification component model overview is also shown. There are other outstanding problems with CBSE that can be looked into more thoroughly.

**References**
[1] I. Crnkovic, J Stafford, C Szyperski," Software Components beyond Programming: From Routines to Services", IEEE Software, pp.22-26 May/June 2011 **DOI:** 10.1109/MS.2011.62
[2] Basha, N.M.J.; Moiz, S.A., "Component based software development: A state of art," Advances in Engineering, Science and Management (ICAESM), 2012 International Conference on , vol., no., pp.599,604, 30-31 March 2012. https://doi.org/10.48550/arXiv.1406.3728
[3] Fahmi, S.A; Ho-Jin Choi, "Life Cycles for Component-Based Software Development," CIT Workshops 2008. IEEE 8th

International Conference on Computer and Information Technology Workshops, pp.637-642, 2008. **DOI:** 10.1109/CIT.2008.Workshops.82
[4] Sommerville I, "Software Engineering", 7th Edition, Pearson Education, 2004. https://dl.acm.org/doi/book/10.5555/983346
[5] I. Crnkovic, S Sentilles, A Vulgarakis, M R.V. Chaudron," A Classification Framework for Software Component Models", IEEE Transactions on Software Engineering Vol. 37 No.5, 2011. **DOI:** 10.1109/TSE.2010.83
[6] Z Chengbang, L Bing, L Shufen," A Component Quality of Service Modeling Method", IEEE 18th International Conference on Computer Supported Cooperative Work in Design, pp.695-699, 2014. **DOI:** 10.1109/CSCWD.2014.6846929
[7] Liu, C., van Dongen, B. F., Assy, N., & van der Aalst, W. M. (2019, May). A General Framework to Identify Software Components from Execution Data. In ENASE (pp. 234-241). https://doi.org/10.5220/0007655902340241
[8] Ajayi, Olusola O., Stella C. Chiemeke, and Kingsley C. Ukaoha. "Comparative analysis of software components reusability level using gfs and ANFIS soft-computing techniques." In 2019 IEEE AFRICON, pp. 1-8. IEEE, 2019. DOI: 10.1109/AFRICON46755.2019.9134021
[9] Garg Rakesh. A ranking model for the selection and ranking of commercial off-the-shelf components. IEEE Transactions on Engineering Management. 2020 Jul 13. DOI: 10.5267/j.dsl.2015.12.004
[10] Basha, N.M.J., Ganapathy, G., Moulana, M. (2022). CREA-Components Reusability Evaluation and Assessment: An Algorithmic Perspective. In: Luhach, A.K., Jat, D.S., Hawari, K.B.G., Gao, XZ., Lingras, P. (eds) Advanced Informatics for Computing Research. ICAICR 2021. Communications in Computer and Information Science, vol 1575. Springer, Cham. https://doi.org/10.1007/978-3-031-09469-9_12